\documentclass[conference]{IEEEtran}
\IEEEoverridecommandlockouts
\usepackage{amsmath,amssymb,amsfonts}
\usepackage[linesnumbered,ruled,vlined,titlenumbered]{algorithm2e}
\usepackage{pgfplots}
\pgfplotsset{compat=newest}
\usepackage{tikz}
\usepackage{tikzscale}
\usepackage{csvsimple} 

\DeclareSymbolFont{EULE}{U}{eur}{m}{n}

\newcommand{\RR}{\ensuremath{\mathbb{R}}}			

\newcommand{\M}[1]{{\mathchoice{\mbox{\boldmath$\displaystyle #1$}}%
		{\mbox{\boldmath$\textstyle #1$}}%
		{\mbox{\boldmath$\scriptstyle #1$}}%
		{\mbox{\boldmath$\scriptscriptstyle #1$}}}} 
\newcommand{\diag}[1]{{\mathchoice{\ensuremath{\mathrm{\mathbf{diag}}\!\left(#1\right)}}
		{\ensuremath{\mathrm{\mathbf{diag}}(#1)}}
		{\ensuremath{\mathrm{\mathbf{diag}}(#1)}}
		{\ensuremath{\mathrm{\mathbf{diag}}(#1)}}
}} 

\newcommand{\MatrId}{\ensuremath{\M{I}}} 
\newcommand{\nullvec}{\ensuremath{\M{0}}} 

\newcommand{\pnorm}[2]{\left\lVert#1\right\rVert_{#2}}
\newcommand{\FrobNorm}[1]{||#1||_\mathrm{F}}

\DeclareMathSymbol{\pdf}{\mathalpha}{EULE}{"66}
\DeclareMathSymbol{\pdq}{\mathalpha}{EULE}{"71}

\newcommand{\rv}[1]{\ensuremath{\mathsf{#1}}} 							
\newcommand{\randvec}[1]{\ensuremath{\M{\mathsf{#1}}}}					

\newcommand{\Gauss}[1]{{\mathchoice{\ensuremath{\mathcal{N}\!\left(#1\right)}}			
			{\ensuremath{\mathcal{N}(#1)}}
			{\ensuremath{\mathcal{N}(#1)}}
			{\ensuremath{\mathcal{N}(#1)}}
}} 

\newcommand{\EW}[2]{{\mathchoice{\ensuremath{\mathrm{E}_{#2}\!\left\{#1\right\}}}
		{\ensuremath{\mathrm{E}_{#2}\{#1\}}}
		{\ensuremath{\mathrm{E}_{#2}\{#1\}}}
		{\ensuremath{\mathrm{E}_{#2}\{#1\}}}	
}} 

\newcommand{\KLdiv}[2]{{\mathchoice{\ensuremath{\operatorname{D}_\mathrm{KL}\!\left(#1\mid\mid#2\right)}}
		{\ensuremath{\operatorname{D}_\mathrm{KL}(#1||#2)}}
		{\ensuremath{\operatorname{D}_\mathrm{KL}(#1||#2)}}
		{\ensuremath{\operatorname{D}_\mathrm{KL}(#1||#2)}}
}} 

\def\argmin{\mathop{\mathrm{argmin}}}			 	
\DeclareMathSymbol{\dirac}{\mathalpha}{EULE}{"0E} 	

\newcommand{\deriv}{\ensuremath{\mathrm{d}}}
\newcommand{\entspr}{\ensuremath{\;\widehat{=}\;}}

\makeatletter
\newcommand{\removelatexerror}{\let\@latex@error\@gobble}
\let\oldnl\nl
\newcommand{\nonl}{\renewcommand{\nl}{\let\nl\oldnl}}
\makeatother

\newcommand{\VAMP}{VAMPavg}
\newcommand{\ECind}{VAMPind} 
\newcommand{\fracVAMP}{frac\VAMP}
\newcommand{\fracECind}{frac\ECind}

\newcommand{\olin}{oLE} 
\newcommand{\onle}{oNLE} 
\newcommand{\cnle}{cNLE} 

\newcommand{\dampfactor}{d}

\newcommand{\fraction}{e}

\def\heightsimrespics{1.8} 

\newcommand{\colidx}{j}
\def\linindex{\mathrm{c}} 
\def\nleindex{\mathrm{s}} 

\newcommand{\itidx}{k}

\newcommand{\permcolidx}{p}
\newcommand{\spreadpp}{v}

\newcommand{\ppmatentry}{r}

\newcommand{\sig}{x}
\newcommand{\sigrv}{\rv{\sig}}
\newcommand{\sigvec}{\ensuremath{\M{\sig}}}
\newcommand{\sigrvec}{\ensuremath{\randvec{\sig}}}

\newcommand{\sigdim}{N}

\newcommand{\sparsity}{s}
\newcommand{\relsparsitycalc}{\mathchoice{\ensuremath{\frac{\sparsity}{\sigdim}}}
	{\ensuremath{\sparsity/\sigdim}}
	{\ensuremath{\sparsity/\sigdim}}
	{\ensuremath{\sparsity/\sigdim}}
}

\newcommand{\obsdim}{\ensuremath{M}}
\newcommand{\yvec}{\M{y}}
\newcommand{\yrvec}{\randvec{y}}

\newcommand{\A}{\ensuremath{\M{A}}} 	

\newcommand{\noise}{n}

\newcommand{\noiservec}{\ensuremath{\randvec{\noise}}}
\newcommand{\noiserv}{\ensuremath{\rv{\noise}}}

\newcommand{\posteriorarg}[1]{\pdf_{\sigrvec\mid\yrvec}(#1)}
\newcommand{\posterior}{\posteriorarg{\sigvec}}
\newcommand{\soenorm}[1]{\pnorm{\yvec-\A#1}{2}^2} 

\newcommand{\factor}{f}
\newcommand{\factorarg}[1]{f_{#1}}
\newcommand{\flin}{\factorarg{\linindex}}
\newcommand{\fnle}{\factorarg{\nleindex}}

\newcommand{\mean}{m}
\newcommand{\meanvec}{\M{\mean}}

\newcommand{\suffstat}{g}

\newcommand{\suffstatvec}{\M{\suffstat}}

\newcommand{\partfunc}{Z}

\newcommand{\natpar}{\theta}
\newcommand{\natparvec}{\M{\natpar}}

\newcommand{\lam}{\lambda}
\newcommand{\Lam}{\Lambda}

\newcommand{\natparveclin}{\natparvec_{\linindex}}
\newcommand{\natparvecnle}{\natparvec_{\nleindex}}
\newcommand{\natparveclinarg}[1]{\natparvec_{\linindex,#1}}
\newcommand{\natparvecnlearg}[1]{\natparvec_{\nleindex,#1}}

\newcommand{\trialdistr}{\pdq}
\newcommand{\cavtrialdistr}[1]{\pdq^{\backslash #1}}

\newcommand{\natparveccavity}[1]{\natparvec^{\backslash#1}}

\newcommand{\Lamcavity}[1]{\Lam^{\backslash#1}}

\newcommand{\sigidx}{\sig_\colidx}

\newcommand{\meanlinvec}{\meanvec_{\linindex}}

\newcommand{\meanlinidx}{\mean_{\linindex,\colidx}}

\newcommand{\meannlevec}{\meanvec_{\nleindex}}

\newcommand{\meannleidx}{\mean_{\nleindex,\colidx}}

\newcommand{\prevestlinidx}{\tilde{\sig}_{\linindex,\colidx}}
\newcommand{\prevestlinvec}{\tilde{\sigvec}_{\linindex}}

\newcommand{\prevestnleidx}{\tilde{\sig}_{\nleindex,\colidx}}

\newcommand{\noisevar}{\ensuremath{\sigma_{\noiserv}^2}} 
\newcommand{\mismatchnoisevar}{\tilde{\sigma}_{\noiserv}^2}

\newcommand{\varx}{\sigma_{\sigrv}^2} 

\newcommand{\varprevestlinarg}[1]{\tilde{\sigma}_{\linindex,#1}^2}
\newcommand{\varprevestnlearg}[1]{\tilde{\sigma}_{\nleindex,#1}^2}
\newcommand{\varprevestlinidx}{\varprevestlinarg{\colidx}}
\newcommand{\varprevestnleidx}{\varprevestnlearg{\colidx}}

\newcommand{\varlinarg}[1]{\sigma_{\linindex,#1}^2}
\newcommand{\varnlearg}[1]{\sigma_{\nleindex,#1}^2}
\newcommand{\varlinidx}{\varlinarg{\colidx}}
\newcommand{\varnleidx}{\varnlearg{\colidx}}

\newcommand{\covar}{\ensuremath{\M{\Phi}}}
\newcommand{\covarprevestlin}{\tilde{\covar}_{\linindex}}

\newcommand{\Lnlearg}[1]{\Lam_{\nleindex,#1}}
\newcommand{\Lnleidx}{\Lnlearg{\colidx}}

\begin{document}

\title{Stabilization Techniques for Iterative Algorithms in Compressed Sensing
\thanks{This work was funded by the Deutsche Forschungsgemeinschaft
	(DFG, German Research Foundation) --- FI 982/16-1.
}
}

\author{
	\IEEEauthorblockN{Carmen Sippel, 
		Robert F.H.~Fischer
	}
	\IEEEauthorblockA{
		Institute of Communications Engineering, Ulm University, Germany \\
		{\tt\small \{carmen.sippel, robert.fischer\}@uni-ulm.de}
	}
}

\maketitle

\begin{abstract}
	Algorithms for signal recovery in compressed sensing (CS) are often improved by stabilization techniques, such as damping, or the less widely known so-called fractional approach, which is based on the expectation propagation (EP) framework. 
	These procedures are used to increase the steady-state performance, i.e., the performance after convergence, or assure convergence, when this is otherwise not possible. 
	In this paper, we give a thorough introduction and interpretation of several stabilization approaches. 
	The effects of the stabilization procedures are examined and compared via numerical simulations and we show that a combination of several procedures can be beneficial for the performance of the algorithm. 
\end{abstract}


\vspace{-0.1cm}
\section{Introduction}
\label{sec:intro}
\noindent
Compressed sensing (CS)~\cite{Candes2006robustUncertainty:SignalReconstrCS,Donoho2006CompressedSensing} deals with an underdetermined system of linear combinations of the transmit symbols in noise, where the signal to be recovered is assumed to be sparse, i.e., has few non-zero entries.
This imposes two constraints on the problem. 
Iterative algorithms for signal recovery in CS generally solve the problem by ignoring one constraint, while solving the other and vice versa. 
The state-of-the-art iterative algorithm is currently, besides \emph{approximate message passing (AMP)}~\cite{Maleki2011AMP}, the so-called \emph{vector approximate message passing (VAMP)}~\cite{Rangan2019VAMP_long}, which can be derived from the expectation propagation (EP) framework~\cite{Minka2001EP}. 

It has been reported that \emph{damping} can increase the steady-state performance of VAMP or even tip the scale in terms of convergence at all, especially in challenging scenarios, e.g., for ill-conditioned measurements~\cite{Rangan2019VAMP_long}. 
Generally, damping is a widely used approach to enable convergence of recovery algorithms in CS, when this is otherwise not given~\cite{Cakmak2014SAMP-GeneralMatrixEnsembles,Kabashima2014SRECLin,Vila2015AdaptiveDampMeanRemovGAMP}. 
From the EP framework another approach, known as \emph{fractional EP} or \emph{power EP}~\cite{Minka2004PowerEP,Minka2005alpha-divergenceMP} is known to improve performance if standard EP fails~\cite{Li2016UnifyingApproxInfFrameworkVarFreeEnergyRelax,Seeger2008BayesInfOptDesignSparseLinModel}. 
For the CS scenario, this has successfully been applied in~\cite{Seeger2008CS_BayesExpDesign}. 

In the literature, a thorough understanding and interpretation of the principles is missing. 
In this paper, we examine the effects of several damping approaches and show that in the CS setting the fractional approach leads to a  procedure similar to damping. 
We give an interpretation of the approaches and show by numerical simulations how performance can be improved by combining the techniques. 

The paper is organized as follows. 
In Sec.~\ref{sec:SysModel} we introduce the system model for CS and state the recovery problem. 
In Sec.~\ref{sec:EP}, VAMP is briefly motivated by the EP framework. 
Section~\ref{sec:StabTech} introduces the different stabilization techniques, which are compared and interpreted in Sec.~\ref{sec:Discussion}. 
We show results of numerical simulations in Sec.~\ref{sec:numRes}, and conclude our work in Sec.~\ref{sec:Conclusion}.

This paper is an extended version of~\cite{Sippel2021FracEC}.

\section{System Model for Compressed Sensing}
\label{sec:SysModel}
\noindent
We model the noisy CS measurements 
by\footnote{We denote scalars by small letters, e.g., $x$, vectors by bold ones,
	e.g., $\M{x}$, matrices by upper case bold, e.g., $\M{X}$,
	and random variables in sans-serif font, i.e., $\rv{x}$, $\randvec{x}$,
	and $\randvec{X}$, respectively.
	$\MatrId_m$: $m\times m$ identity matrix, 
	$\nullvec$: all-zero vector, 
	$\diag{\cdot}$: diagonal matrix with given entries, 
	$\EW{\cdot}{\rv{x},\pdf_{\rv{x}}}$: expectation of random variable $\rv{x}$ w.r.t.\ pdf $\pdf_{\rv{x}}(x)$, 
	$\pi(\cdot)$: random permutation, 
	$[\cdot]_j$: $j$th entry, 
	$\ln(\cdot)$: natural logarithm.
} 
\begin{align} 					\label{eq:CS_SysModel}
\yrvec &= \A \sigrvec + \noiservec \in \RR^\obsdim\;,
\end{align}
where the sensing matrix $\A \in \RR^{\obsdim \times \sigdim}$, $\obsdim \ll \sigdim$, is assumed to be known and the noise
$\noiservec \sim \Gauss{\nullvec,\noisevar \MatrId_\obsdim}$ is i.i.d.\ Gaussian. The elements $\sigrv_\colidx$ of
$\sigrvec = [\sigrv_1,\dots,\sigrv_\sigdim]^\top$ are assumed to be i.i.d.\ with marginal probability density function (pdf)
$\pdf_\sigrv(\sigidx)$, 
i.e., 
\begin{align} 						\label{eq:pdf_x_separable}
\pdf_{\sigrvec}(\sigvec) = \prod\nolimits_{\colidx=1}^{\sigdim} \pdf_{\sigrv}(x_\colidx) \;.
\end{align}
We assume a sparse signal $\sigrvec$, which is reflected by a Dirac delta function at zero in the marginal pdf. 
Hence, the overall problem is specified by the posterior
\begin{align}			\label{eq:posterior_Bayes}
\posterior = \frac{1}{\pdf_{\yrvec}(\yvec)} \pdf_{\sigrvec}(\sigvec) \cdot \pdf_{\yrvec\mid\sigrvec}(\yvec \mid \sigvec) \; ,
\end{align}
where 
$
\pdf_{\yrvec\mid\sigrvec}(\yvec \mid \sigvec) = 
\frac{1}{\sqrt{(2\pi \noisevar)^{\obsdim}}} \exp\!\left( -\frac{1}{2\noisevar} \soenorm{\sigvec} \right) 
$ 
due to the additive Gaussian noise $\noiservec$. 
We omit the argument $\yvec$ in the posterior distribution for brevity. 
The task of recovering the signal $\sigrvec$ is given by the estimation problem
\begin{align}					\label{eq:est_CS}
\EW{\sigrvec \mid \yvec,\, \A,\,  \noisevar}{\sigrvec,\pdf_{\sigrvec\mid\yrvec}} &= \int \sigvec \, \posterior \, \deriv \sigvec \; ,
\end{align}
which is infeasible for high-dimensional $\sigvec$.
Hence, there is a need for suitable algorithms. 

\section{Derivation of VAMP based on EP}\label{sec:EP}
\noindent
The main idea of \emph{expectation propagation} (EP)~\cite{Minka2001EP} is to simplify the estimation~\eqref{eq:est_CS} by approximating the distribution $\posterior$ (globally) by iteratively simplifying parts of it (locally) and \emph{projecting}~\cite{Minka2005alpha-divergenceMP} this simplification onto the global approximation, which  
we call $\trialdistr(\sigvec)$. 
The projection is motivated by the use of \emph{exponential families}~\cite{Brown1986fundamentalsStatExpFamDecisionTheory} in combination with the Kullback-Leibler divergence
\begin{align}
\KLdiv{\posterior}{\trialdistr(\sigvec)} = \int \posterior \ln \frac{\posterior}{\trialdistr(\sigvec)} \, \deriv \sigvec \; ,
\end{align}
which causes a so-called \emph{matching of moments}~\cite{Opper2005ecai}.

\subsection{Exponential Families}
\label{sec:ExpFam}
\noindent
An exponential family is parameterized by the so-called \emph{natural parameters} $\natparvec$ and given by~\cite{Brown1986fundamentalsStatExpFamDecisionTheory}
\begin{align}
\trialdistr(\sigvec) &= \frac{1}{\partfunc(\natparvec)} \exp\!\left( \natparvec^\top \suffstatvec(\sigvec) \right) \; ,
\end{align}
where $\partfunc(\natparvec)= \int \exp( \natparvec^\top \suffstatvec(\sigvec) ) \, \deriv \sigvec$ serves as normalization. 
When specifying the exponential family by the first two moments, i.e., defining the \emph{sufficient statistics} as $\suffstatvec(x) = [ x, \, -x^2/2 ]^\top$ (with corresponding $\natparvec = [\lam,\,\Lam]^\top$) the family of Gaussian distributions is obtained. 
The connection between natural and moment parameters (mean $\mean$ and variance $\sigma^2$) is then given by
\begin{align} 						\label{eq:connection_mom_nat_par_Gauss}
\lam = \frac{\mean}{\sigma^2} \; , \qquad \Lam = \frac{1}{\sigma^2} \; .
\end{align}

The projection onto an exponential family, inherently causes a matching of moments~\cite{Opper2005ecai}.
We explain the matching of moments for the given problem below. 

\subsection{Structure of the Problem}
\noindent
The structure~\eqref{eq:posterior_Bayes}, shows a factorization into two parts (that depend on $\sigvec$). 
We denote the part based on signal prior $\pdf_{\sigrvec}(\sigvec)$ by ``$\nleindex$'' for \emph{signal constraint} and the second one by ``$\linindex$'' for \emph{channel constraint}, i.e.,
\begin{align} 					\label{eq:factorization_EC}
\posterior = \frac{1}{\partfunc} \fnle(\sigvec) \cdot \flin(\sigvec) \; .
\end{align}
The (local) approximations are then obtained by replacing either of the factors by an \emph{exponential family}~\cite{Brown1986fundamentalsStatExpFamDecisionTheory}, i.e., we define the \emph{tilted} exponential family~\cite{Seeger2008BayesInfOptDesignSparseLinModel}
$	\cavtrialdistr{\bullet}(\sigvec) 
= \frac{1}{\partfunc_{\bullet}(\natparveccavity{\bullet})} \factor_{\bullet}(\sigvec) \exp( (\natparveccavity{\bullet})^\top \suffstatvec(\sigvec) ) $  for 
$\bullet \in \{\linindex,\,\nleindex\} 
$. 
We choose this notation for brevity, since $\natparveccavity{\bullet}$ captures the moments obtained from the other (complementary) constraint.
Considering the overall approximation $\trialdistr(\sigvec)$, we obtain these approximations by first removing the part of the approximation that represents the respective factor (\emph{exclusion}) and then inserting the actual factor $\factor_{\bullet}(\sigvec)$ for $\bullet \in \{\linindex,\,\nleindex\}$ (\emph{inclusion}). 
Note that, since we consider two factors, the EP approach coincides with expectation-consistent (EC) approximate inference~\cite{Opper2005ecai}. 

\subsection{Procedure}
\noindent
Either of the parts in~\eqref{eq:factorization_EC} is iteratively replaced by a pdf from an exponential family, followed by the respective projection~\cite{Minka2005alpha-divergenceMP}
\begin{align} 								\label{eq:proj}
\trialdistr_{\bullet}(\sigvec) &= \argmin_{\trialdistr(\sigvec)} \KLdiv{ \cavtrialdistr{\bullet}(\sigvec) }{\trialdistr(\sigvec)} \; , \;\; \bullet \in \{\linindex,\,\nleindex\} \; .
\end{align} 
Eq.~\eqref{eq:proj} is solved by the \emph{matchings of moments}~\cite{Opper2005ecai}, which reads
\begin{align} 							\label{eq:mom_match}
\EW{\suffstatvec(\sigrvec)}{\sigrvec,\cavtrialdistr{\bullet}} &= \EW{\suffstatvec(\sigvec)}{\sigrvec,\trialdistr_{\bullet}} \; \, \quad \bullet \in \{\linindex,\,\nleindex\} \; .
\end{align}
This way, the infeasible computation~\eqref{eq:est_CS} is replaced by two feasible ones (left hand-side of~\eqref{eq:mom_match}). 

The exchange between the parameters of both subproblems is justified by considering \emph{exclusions} and \emph{inclusions} from $\trialdistr(\sigvec)$. 
In terms of natural parameters inclusion and exclusion transform to simple additions and subtractions. 
This means, for the given structure with only two factors, the update between the overall approximation, represented by $\natparvec$ and the partial approximations is given by the connection~\cite{Opper2005ecai} $\natparvec = \natparveccavity{\linindex} + \natparveccavity{\nleindex}$. 
In order to distinguish the natural parameters obtained from the different projections, we may resort to using indices, i.e., $\natparvecnle$ and $\natparveclin$, respectively, instead of $\natparvec$, i.e., 
\begin{align}							\label{eq:conn_natpar_EC}
\natparveccavity{\linindex} = \natparvecnle - \natparveccavity{\nleindex} \; , \quad \natparveccavity{\nleindex} = \natparveclin - \natparveccavity{\linindex} \; .
\end{align}%

All in all, the procedure is given by computing one of the left hand-side expectations in~\eqref{eq:mom_match}, mapping from the moment parameters to the natural parameters, e.g., via~\eqref{eq:connection_mom_nat_par_Gauss} to obtain $\natparveclin$, or $\natparvecnle$, respectively, and compute the parameters for the other expectation by~\eqref{eq:conn_natpar_EC}. 

\subsection{Expectations}
\noindent
We stick to Gaussian distributions, when using exponential families, i.e., we define for $\colidx \in \{1,\,\dots,\,\sigdim\}$, $\bullet \in \{\linindex,\,\nleindex\}$:
\begin{align} 							\label{eq:spec_gj_gauss_ind}
\suffstatvec_{\colidx}(\sigidx) &= [\sigidx, - \sigidx^2 / 2]^\top \; ,
\\ 										\label{eq:spec_natpar_gauss_ind}
\natparvec_{\colidx} &= [\lam_{\colidx} , \;\;\;\;\;\; \Lam_{\colidx} ]^\top = [\mean_{\colidx} / \sigma_{\colidx}^2 , \, 1 / \sigma_{\colidx}^2
]^\top \; .
\end{align}
For this case, it is useful to consider all natural parameters as a $2\times\sigdim$-matrix, e.g.,  $\natparvec = [\natparvec_{1} , \, \dots, \, \natparvec_{\sigdim}]$. 
Keeping the channel-constrained part, while replacing the prior, yields a joint linear estimator (LE) given by ($\covarprevestlin = \diag{\varprevestlinidx}$)
\begin{align}			\label{eq:LMMSE}
\meanlinvec 
&=\prevestlinvec + \left( \A^\top \A + \noisevar \covarprevestlin^{-1} \right)^{-1} \A^\top (\yvec - \A \prevestlinvec) \; ,
\\
\label{eq:condvar_LMMSE}
\covar_{\linindex} &= \noisevar \left( \A^\top \A + \noisevar \covarprevestlin^{-1} \right)^{-1}
\; , \quad \varlinidx = [\covar_{\linindex}]_{\colidx\colidx}
\; .
\end{align}

Since the signal prior~\eqref{eq:pdf_x_separable} is separable, the estimation for the signal-constrained part can be calculated individually for each variable $\sigidx$ ($\colidx \in \{1,\,\dots,\,\sigdim\}$).
Hence, we obtain individual, non-linear estimators (NLEs)
\begin{align}							\label{eq:nle_mean}
\textstyle
\meannleidx 
&= 
{\textstyle \frac{1}{\partfunc_{\nleindex,\colidx}}} \int \sig \, \pdf_{\sigrv}(\sig) \exp\!\left( \textstyle \frac{\prevestnleidx}{\varprevestnleidx} \sig - \frac{\sig^2}{2\varprevestnleidx} \right) \, \deriv \sig \; ,
\\										\label{eq:nle_condvar}
\varnleidx &= {\textstyle \frac{1}{\partfunc_{\nleindex,\colidx}}} \!\! \int \! (\sig \! - \! \meannleidx)^2 \pdf_{\sigrv}(\sig) \exp\!\left( \textstyle \frac{\prevestnleidx}{\varprevestnleidx} \sig \!-\! \frac{\sig^2}{2\varprevestnleidx} \right) \deriv \sig ,
\end{align}
with $\partfunc_{\nleindex,\colidx} = \int \pdf_{\sigrv}(\sig) \exp( \textstyle \frac{\prevestnleidx}{\varprevestnleidx} \sig - \frac{\sig^2}{2\varprevestnleidx} ) \, \deriv \sig$. 

\subsection{VAMP}
\label{sec:VAMP}
\noindent
Employing the expectations above into the EP framework, yields a form of the so-called \emph{vector approximate message passing} (VAMP)~\cite{Rangan2019VAMP_long} algorithm with individual variances; which was already introduced in~\cite{Fletcher2016gec}.
The standard approach as given in~\cite{Rangan2019VAMP_long} uses average variances 
\begin{align} 					
\label{eq:avg_var}
\sigma_{\bullet}^2 &= \frac{1}{\sigdim} \sum\nolimits_{\colidx=1}^{\sigdim} \sigma_{\bullet,\colidx}^2 \; , \quad \bullet \in \{\linindex,\,\nleindex\} \; .
\end{align}
For the connection to the moments specified in $\suffstatvec(\sigvec)$, see~\cite{Sippel2020InvarianceVarSepEC}.
The updates between the estimations are then performed with averaged values of the variances, instead of the individual ones. 
We will call the version with average and individual variances, \VAMP\ and \ECind, respectively. 

\section{Stabilization Techniques}
\label{sec:StabTech}
\noindent
The introduced procedures are very powerful recovery algorithms. 
However, for \ECind\ numerical issues with the variances have been reported~\cite{Fischer2019VAMPIndividual}, causing a degradation in performance. 
Furthermore, also \VAMP\ underlies a drop in performance, when a non-uniform power distribution over the components of $\sigvec$ is present~\cite{Sippel2021seqVAMPire}. 
We examine two strategies for the stabilization of these algorithms. 
So-called \emph{damping} is widely used in the CS literature, whenever algorithms have problems to converge~\cite{Cakmak2014SAMP-GeneralMatrixEnsembles,Kabashima2014SRECLin,Rangan2019VAMP_long}. 
The second approach is known as \emph{fractional approach}~\cite{Wiegerinck2003FractionalBP} or \emph{power EP}~\cite{Minka2004PowerEP} and closely related to damping, as we will show in the following.  

\subsection{Damping Procedures}
\label{sec:Damping}
\noindent
Damping is based on convex combination of previous and current estimates. 
There are several possibilities for such combinations.
We consider the following ones, using $\dampfactor \in (0,\,1]$ as damping parameter ($\dampfactor=1$ means no damping) and $\itidx$ as iteration index
\begin{itemize}\setlength{\itemindent}{2em}
	\item[\onle)] $\natparvecnle = \dampfactor \natparvecnle + (1-\dampfactor) \natparveclin$
	\item[\olin)] $\natparveclin = \dampfactor \natparveclin + (1-\dampfactor) \natparvecnle$
	\item[\cnle)] $\natparvecnle^{[\itidx]} = \dampfactor \natparvecnle^{[\itidx]} + (1-\dampfactor) \natparvecnle^{[\itidx-1]} $
\end{itemize}
The first two strategies, \onle\ and \olin, oppose each other; one is applied after the LE, one after the NLE.
The effect can be thought of as partly (depending on $\dampfactor$) removing the effects of one of the estimations and replacing it by the other one. 
\onle\ \emph{\underline{o}mits} (for $\dampfactor \to 0$) the NLE, \olin\ omits the LE.
The third strategy \cnle\ is often used in the literature, e.g., in~\cite{Rangan2019VAMP_long}, and also partly removes the current estimates but sticks to the NLEs---it \emph{\underline{c}ombines} the NLEs of successive iterations. 
The corresponding opposing strategy, which would combine the linear estimates, does not yield good results, because the linear estimate is usually the less reliable one. 
The procedure is therefore not considered here. 

\subsection{Fractional Approach}
\label{sec:fracEC}
\noindent
The idea of the fractional approach is to introduce a parameter, we call it $\fraction$, that enables to \emph{partly} remove and insert the approximation and factor parts from and to~\eqref{eq:factorization_EC}, respectively.
Since the expectations for the signal-constrained part cannot be straightforwardly computed for arbitrary $\fraction$, we consider only the case of partially removing and inserting the channel-constrained part $\flin(\sigvec)$. 
The parameter $\fraction$ is applied such that $1/\fraction$ specifies by what \emph{fraction} (in the domain of natural parameters) the current approximation of $\flin(\sigvec)$ is removed in order to be replaced by a respective fraction of the actual factor. 
In particular, the tilted exponential family becomes
$\cavtrialdistr{\linindex}_{\fraction}(\sigvec) = \frac{1}{\partfunc_{\linindex}(\natparveccavity{\linindex})} \flin^{1/\fraction}(\sigvec) \exp( (\natparveccavity{\linindex})^\top \suffstatvec(\sigvec) )$. 
The updates for the natural parameters turn to~\cite{Seeger2005EPExpFam}
\begin{align} 					\label{eq:conn_natpar_fractional}
\natparveccavity{\linindex} = \natparvecnle - \frac{1}{\fraction} \natparveccavity{\nleindex} \; , \quad 
\natparveccavity{\nleindex} = \fraction(\natparveclin - \natparveccavity{\nleindex}) \; .
\end{align}
The respective expectations become
\begin{align}
\meanlinvec 
&= {\textstyle \frac{1}{\partfunc_{\linindex}(\natparveccavity{\linindex})} } \!\! \int \!\! \sigvec \flin^{1/\fraction}(\sigvec) \exp\!\left( \prevestlinvec^\top \covarprevestlin^{-1} \sigvec - \frac{1}{2} \sigvec^\top \covarprevestlin^{-1} \sigvec \right) \deriv \sigvec
\nonumber\\ 				
\label{eq:LMMSE_fractional}
&=\prevestlinvec + \left( \A^\top \A + \fraction \noisevar \covarprevestlin^{-1} \right)^{-1} \A^\top (\yvec - \A \prevestlinvec) \; ,
\\ 								\label{eq:condvar_LMMSE_fractional}
\covar_{\linindex} &= \fraction \noisevar \left( \A^\top \A + \fraction \noisevar \covarprevestlin^{-1} \right)^{-1}
\end{align}

\subsection{Clipping}
\noindent
Another standard procedure for numerical stability is clipping~\cite{Rangan2019VAMP_long}, i.e., bounding the range of possible values for respective parameters. 
This needs to be performed for the variances, respectively precisions (inverse variances), in order to keep useful values. 
Especially negative variances are not reasonable and must therefore be forbidden. 
In the following, we do not examine the effect of different clipping bounds. 
Instead, we restrict to the following procedure, which turned out to be relatively stable for a wide range of parameters; we clip $\Lnleidx = 1 / \varnleidx$ to the interval $[10^{-8},\; 10^8]$ and other precisions to $[10^{-12},\; 10^{12}]$. 

\subsection{Complete Algorithm}
\noindent
We combine the stabilization techniques in the following algorithm, which we state for the individual variances case. 
Due to the generalization by the fractional approach, we call it \fracECind. 
Note that the natural parameters are considered to be $2\times\sigdim$-matrices, analogously to above we define $\natparvec_{\bullet} = [\natparvec_{\bullet,1},\,\dots,\,\natparvec_{\bullet,\sigdim}]$ with $\bullet \in \{\linindex,\,\nleindex\}$. 
We do not explicit state clipping here; in the simulations, the procedure described above is used. 
The variant with average variances, \fracVAMP, needs additionally the calculations in~\eqref{eq:avg_var}. 
	
\begin{algorithm}
	\footnotesize
	\renewcommand*\baselinestretch{0.75}%
	\DontPrintSemicolon
	$\covarprevestlin = \varx \MatrId_{\sigdim}$, \ 
	$\prevestlinvec = \nullvec$, \ 
	$\natparvecnle = [\nullvec , \, [1/\varx, \, \dots, \, 1/\varx ]^\top  ]^\top \in \RR^{2\times\sigdim}
	$ \;
	\While{stopping criterion not met}{
		$\meanlinvec = \prevestlinvec + ( \A^\top \A + \fraction \noisevar \covarprevestlin^{-1} )^{-1} \A^\top (\yvec - \A \prevestlinvec)$\;
		$\covar_{\linindex} = \fraction \noisevar ( \A^\top \A + \fraction \noisevar \covarprevestlin^{-1} )^{-1} $\;
		$\natparveclinarg{\colidx} = [\meanlinidx / [\covar_{\linindex}]_{\colidx\colidx}, \, 1 / [\covar_{\linindex}]_{\colidx\colidx}]^\top \quad \forall \colidx \in \{1,\,\dots,\sigdim\}$ \;
		\If{damping case $=$ \olin\tcp*{Damping Case \olin}}{
			$\natparveclin = \dampfactor \natparveclin + (1-\dampfactor) \natparvecnle$ 
		}
		$\natparveccavity{\nleindex} = \fraction (\natparveclin - \natparveccavity{\linindex})$ \tcp*{Parameter Update~\eqref{eq:conn_natpar_fractional}}
		\nonl\;
		$\meannlevec = \EW{\sigrvec }{\sigrvec,\cavtrialdistr{\nleindex}}$\tcp*{NLE, cf.~\eqref{eq:nle_mean},~\eqref{eq:nle_condvar}}
		$\varnleidx = \EW{ (\sigrv - \meannleidx)^2 }{\sigrv,\cavtrialdistr{\nleindex}_{\colidx}}  \quad \forall \colidx \in \{1,\,\dots,\sigdim\}$\;
		$\natparvecnlearg{\mathrm{old}} = \natparvecnle$\;
		$\natparvecnlearg{\colidx} = [\meannleidx / \varnleidx, \, 1 / \varnleidx]^\top \quad \forall \colidx \in \{1,\,\dots,\sigdim\}$ \;
		\Switch{damping case}{
			\Case{\onle\tcp*{Damping Case \onle}}{
				$\natparvecnle = \dampfactor \natparvecnle + (1-\dampfactor) \natparveclin$ 
			}
			\Case{\cnle\tcp*{Damping Case \cnle}}{
				$\natparvecnle = \dampfactor \natparvecnle + (1-\dampfactor) \natparvecnlearg{\mathrm{old}}$ 
			}
		}
		$\natparveccavity{\linindex} = \natparvecnle - \natparveccavity{\nleindex} / \fraction$ \tcp*{Parameter Update~\eqref{eq:conn_natpar_fractional}}
		$\prevestlinidx = [\natparveccavity{\linindex}_{\colidx}]_{1} / [\natparveccavity{\linindex}_{\colidx}]_{2} \quad \forall \colidx \in \{1,\,\dots,\sigdim\}$ \;
		$\covarprevestlin = \diag{ 1/[\natparveccavity{\linindex}_{1}]_{2} ,\,\dots,\, 1/[\natparveccavity{\linindex}_{\sigdim}]_{2} }$ \;
	}
	\caption{\mbox{$\meannlevec
			= \mathtt{\fracECind}(\yvec, \, \A,\, \noisevar, \, \varx, \, \fraction, \, \dampfactor)$} 
	}
	\label{alg:fracVAMPind}
\end{algorithm}

\section{Discussion}
\label{sec:Discussion}

\subsection{Estimation Theoretic Interpretation}
\noindent
The expectations~\eqref{eq:LMMSE},~\eqref{eq:condvar_LMMSE} for the EP approach under a Gaussian assumption are identical to the linear minimum mean-squared error (LMMSE) solution, i.e., conditional expectations. 
This means, we estimate the signal $\sigvec$ from an observation $\yvec$ in Gaussian noise, given by the channel~\eqref{eq:CS_SysModel}, where we assume that $\sigrvec$ is Gaussian distributed. 
This assumption might be far from reality, if, e.g., discrete priors as in~\cite{Fischer2019VAMPIndividual,Sippel2021seqVAMPire} are used. 
The fractional approach can be seen as a way to compensate for that inaccuracy. 
As we see in the expectations~\eqref{eq:LMMSE_fractional} and~\eqref{eq:condvar_LMMSE_fractional}, the main change to~\eqref{eq:LMMSE} and~\eqref{eq:condvar_LMMSE} is that we have an effective noise variance $\mismatchnoisevar = \fraction \noisevar$.
Hence, for $\fraction=1$ the standard case is recovered. 
For $\fraction > 1$, we assume a larger noise variance than actually given, the estimate might hence be less sure about its estimation, which gives room for corrections. 
For $\fraction < 1$, the opposite behavior is given, i.e., a lower noise variance than actually given is assumed. 

\subsection{Comparison of Damping and Fractional Approach}
\label{sec:CompDampFrac}
\noindent
Combining the two fractional updates from~\eqref{eq:conn_natpar_fractional} (using iteration index $\itidx$), we obtain
\begin{align}
\natparveccavity{\nleindex,[\itidx+1]} &= \fraction ( \natparveclin^{[\itidx+1]} - \natparvecnle^{[\itidx]} + \frac{1}{\fraction} \natparveccavity{\nleindex,[\itidx]}) 
\nonumber \\
&=  \fraction ( \natparveclin^{[\itidx+1]} - \natparvecnle^{[\itidx]}) + \natparveccavity{\nleindex,[\itidx]}\; .
\end{align}
This resembles very closely the computation, obtained from combining the parameter update~\eqref{eq:conn_natpar_EC} with damping using case~\olin\ for $\fraction=1$
\begin{align}
\natparveccavity{\nleindex,[\itidx+1]} &= \dampfactor \natparveclin^{[\itidx+1]} + (1-\dampfactor) \natparvecnle^{[\itidx]} - \natparveccavity{\linindex,[\itidx]} 
\nonumber \\
&
= \dampfactor ( \natparveclin^{[\itidx+1]} - \natparvecnle^{[\itidx]}) + \natparveccavity{\nleindex,[\itidx]} \; .
\end{align}
Both parameters, $\dampfactor$ and $\fraction$, thus cause the same update here. 
The difference is that in the fractional approach the estimations have a different effective noise variance $\mismatchnoisevar = \fraction \noisevar$ and the $\fraction$ is also considered in the second update derived from~\eqref{eq:conn_natpar_fractional}. 

\subsection{Numerical Considerations}
\label{sec:NumConsider_EC_negVar}
\noindent
The main problem in \ECind, that causes stability problems is the fact that the subtraction~\eqref{eq:conn_natpar_EC} of the precision parameter (see~\eqref{eq:spec_natpar_gauss_ind}) $\Lamcavity{\linindex}_{\colidx} = \Lnleidx - \Lamcavity{\nleindex}_{\colidx}$ turns frequently negative, resulting in negative (extrinsic) variances, which have no reasonable meaning. 
Even with clipping, a degradation in performance may be visible in comparison to the average case \VAMP\ for certain scenarios~\cite{Fischer2019VAMPIndividual}. 
The fractional version stated above, adjusts the update to $\Lamcavity{\linindex}_{\colidx} = \Lnleidx - \Lamcavity{\nleindex}_{\colidx} / \fraction$. 
For $\fraction\uparrow$, we expect the performance of \fracECind\ to increase, since the subtrahend and therefore the probability of a negative result decreases.

\section{Numerical Results}
\label{sec:numRes}
\noindent
We examine the behavior of the above introduced algorithms over the parameters $\dampfactor$ and $\fraction$.

As prior pdf, we use the Bernoulli-Gaussian distribution with sparsity $\sparsity=12$ and variance $\relsparsitycalc = 0.0465$ and Dirac delta function $\dirac(\cdot)$ given by
\begin{align}								\label{eq:pdf_x_DCS}
\textstyle
\pdf_\rv{x}(x) = (1-\relsparsitycalc) \dirac(x) + \frac{\sparsity}{\sigdim} \frac{1}{\sqrt{2\pi}} \exp(-x^2 / 2)\; .
\end{align}
The $129\times258$-sensing matrix $\A$ is i.i.d.\ Gaussian distributed. 
We consider a non-uniform power distribution for the transmit powers of signal $\sigvec$ and amplify the columns of $\A$ accordingly. 
The power distribution is described by the scaling $\ppmatentry_\permcolidx = \spreadpp^{(\permcolidx-1)/(\sigdim-1)}$ ($\permcolidx \in \{1,\,\dots,\,\sigdim\}$); the assignment to the $\colidx$th column is obtained from a random permutation $\colidx = \pi(\permcolidx)$. 
The simulations are obtained for a factor $\spreadpp = 0.2$. 
After applying the power profile (by $\A\cdot \diag{\ppmatentry_{\pi(\permcolidx)}}$), the sensing matrix is scaled such that it has Frobenius norm $\FrobNorm{\A} = \sqrt{\sigdim}$. 
The average condition number of the obtained matrices is $\kappa_{\mathrm{avg}} = 6.33$, for the simulations the condition number ranges in $\kappa \in [5.46,\,7.83]$. 
The signal-to-noise ratio is fixed to $-10\log_{10}(\noisevar) \entspr 17\; \mathrm{dB}$ in all simulations. 

Since the sparsity $\sparsity$ is assumed to be known, we utilize the knowledge and set the smallest $\sigdim - \sparsity$ values in $\meannlevec$ to $0$, cf.~\cite{Sparrer2017IMS&TSR}, before evaluating the performance. 
As performance measure we average the per-symbol normalized mean-squared error (NMSE), $\mathrm{NMSE} = \pnorm{\sigvec - \meannlevec}{2}^2 / \pnorm{\sigvec}{2}^2 / \sigdim$. 
We evaluate the performance after 20 iterations. 

\begin{figure}
	\centering
	\includegraphics[width=0.9\linewidth]{Sim_fracVAMP_SimfracECind_Sim_L258_K129_s12_EbNodB17_It20_priorBG_varp1_pmNMSE_seed2020_d0.2_1_dparonle_SVD0_normA0_varR0.2_epsl1e-08_epsu1e-08_rnggenThreefry_numSensing100000_IterEval20_twocol1.tikz}
	\caption{Performance over damping parameter $\dampfactor$ for damping case \onle, $\sigdim=258$, $\obsdim=129$, $\sparsity=12$, evaluated after 20 iterations.}
	\label{fig:Sim_fracVAMP_SimfracECind_Sim_L258_K129_s12_EbNodB17_It20_priorBG_varp1_pmNMSE_seed2020_d0.2_1_dparonle_SVD0_normA0_varR0.2_epsl1e-08_epsu1e-08_rnggenThreefry_numSensing100000_IterEval20_twocol1}
\end{figure}
Figure~\ref{fig:Sim_fracVAMP_SimfracECind_Sim_L258_K129_s12_EbNodB17_It20_priorBG_varp1_pmNMSE_seed2020_d0.2_1_dparonle_SVD0_normA0_varR0.2_epsl1e-08_epsu1e-08_rnggenThreefry_numSensing100000_IterEval20_twocol1} examines the behavior of both introduced algorithms over the damping parameter $\dampfactor$ for the damping version \onle. 
Especially \VAMP\ (\fracVAMP\ with $\fraction=1$) profits from the damping. 
Furthermore, the performance can be increased by additionally using the fractional approach. 
In the case of \fracVAMP, $\fraction < 1$ leads to an improvement, whereas \fracECind\ requires $\fraction > 1$.
The last part confirms the conjecture from Sec.~\ref{sec:NumConsider_EC_negVar} that \fracECind\ benefits from decreasing the probability of negative variances. 

\begin{figure}
	\centering
	\includegraphics[width=0.9\linewidth]{Sim_fracVAMP_SimfracECind_Sim_L258_K129_s12_EbNodB17_It20_priorBG_varp1_pmNMSE_seed2020_d0.2_1_dparolin_SVD0_normA0_varR0.2_epsl1e-08_epsu1e-08_rnggenThreefry_numSensing100000_IterEval20_twocol1.tikz}
	\caption{Performance over damping parameter $\dampfactor$ for damping case \olin, $\sigdim=258$, $\obsdim=129$, $\sparsity=12$, evaluated after 20 iterations. For the legend of the colored curves, see Fig.~\ref{fig:Sim_fracVAMP_SimfracECind_Sim_L258_K129_s12_EbNodB17_It20_priorBG_varp1_pmNMSE_seed2020_d0.2_1_dparonle_SVD0_normA0_varR0.2_epsl1e-08_epsu1e-08_rnggenThreefry_numSensing100000_IterEval20_twocol1}. }
	\label{fig:Sim_fracVAMP_SimfracECind_Sim_L258_K129_s12_EbNodB17_It20_priorBG_varp1_pmNMSE_seed2020_d0.2_1_dparolin_SVD0_normA0_varR0.2_epsl1e-08_epsu1e-08_rnggenThreefry_numSensing100000_IterEval20_twocol1}
\end{figure}
In Fig.~\ref{fig:Sim_fracVAMP_SimfracECind_Sim_L258_K129_s12_EbNodB17_It20_priorBG_varp1_pmNMSE_seed2020_d0.2_1_dparolin_SVD0_normA0_varR0.2_epsl1e-08_epsu1e-08_rnggenThreefry_numSensing100000_IterEval20_twocol1}, the damping variant \olin\ is considered. 
Here, we additionally plot $\fraction=\dampfactor$, which gives us the chance to compare the cases of a pure fractional approach (triangles) with the respective pure damping procedure (circles), as was theoretically compared in Sec.~\ref{sec:CompDampFrac}. 
Although sharing a common update the behavior of both strategies differs significantly, which can be explained by the fact that the fractional approach also adjusts the estimations. 

Apart from that, we can see that \fracECind\ is more sensitive to changes in $\dampfactor$ than for case \onle, especially for small $\fraction$, which is due to the similarity of the updates in this case, i.e., the effect of damping and fractional approach superimpose. 

The third damping version \cnle\ is not shown here, because it behaves similarly to the first two cases (for \fracECind\ the behavior is close to the case \onle, for \fracVAMP\ it is close to \olin). 

\begin{table}
	\caption{Minima in NMSE over grid of $\dampfactor \in [0.2,\,1]$, $\fraction \in [0.6,\,3.5]$ for the different damping cases.}
	\label{tab:MinimaNMSEgrid}
	\centering
	\resizebox{0.9\linewidth}{!}{%
		\begin{tabular}{lccccccc} 
			\hline
			\hline\\[-2ex]
			&\multicolumn{3}{c}{\fracVAMP} & & \multicolumn{3}{c}{\fracECind} \\
			\cline{2-4} \cline{6-8}\\[-2ex]
			& $\dampfactor$ & $\fraction$ & NMSE & & $\dampfactor$ & $\fraction$ & NMSE \\
			\hline\vspace{-1ex}
			\begin{filecontents*}{Minima.csv}
,\dampfactor,\fraction,NMSE,\dampfactor,\fraction,NMSE
---,1,1,1.2229e-03,1,1,3.9304e-04
\onle,0.51,0.65,3.9894e-04,0.58,3.31,3.5036e-04
\olin,0.55,0.65,3.9970e-04,0.39,3.34,3.4939e-04
\cnle,0.53,0.64,3.9759e-04,0.62,3.30,3.5025e-04
\end{filecontents*}
\csvreader[head to column names]{Minima.csv}{}%
			{\\\vspace{0.5ex}\csvcoli&\csvcolii&\csvcoliii&\csvcoliv&&\csvcolv&\csvcolvi&\csvcolvii}
			\\\hline\hline
		\end{tabular}
	}
\end{table}
In order to get a complete picture, a grid over $\dampfactor \in [0.2,\,1]$ and $\fraction \in [0.6,\,3.5]$ was simulated. 
The resulting minima in NMSE within this area are given in Table~\ref{tab:MinimaNMSEgrid} and compared to the non-stabilized case ($\dampfactor=\fraction=1$ in the first row). 
\begin{figure}
	\centering
	\includegraphics[width=0.9\linewidth]{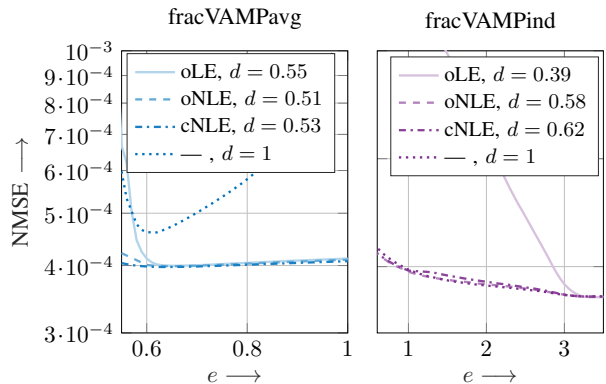}
	\caption{Performance over fractional parameter $\fraction$ for $\sigdim=258$, $\obsdim=129$, $\sparsity=12$, evaluated after 20 iterations.}
	\label{fig:fraclin_Sim}
\end{figure}
The behavior over the fractional parameter $\fraction$ is shown in Fig.~\ref{fig:fraclin_Sim}. 

This example shows that the steady-state performance can be increased by factor 2-3 in terms of NMSE.

\vspace{-0.5ex}
\section{Conclusion}
\label{sec:Conclusion}
\noindent
This work considered stabilization techniques for iterative algorithms in compressed sensing.
The procedures of damping and fractional updates were examined and interpreted in detail. 
In numerical simulations it was shown that a combination of both strategies can supersede each of the stabilization approaches as well as the standard procedure. 

\bibliographystyle{IEEEtranS}
\bibliography{IEEEabrv.bib,refs.bib}

\end{document}